\documentclass[preprint,a4paper,aps,superscriptaddress]{revtex4-1}
\usepackage[english]{babel}
\usepackage[utf8]{inputenc}
\usepackage{mathtools}
\usepackage{enumerate}
\usepackage{amsfonts}
\usepackage{amsthm}
\usepackage{amsmath}
\usepackage{amssymb}
\usepackage{graphicx}
\usepackage{hyperref}
\usepackage{mathrsfs}
\usepackage{bm}

\makeatletter
\newcommand*{\centerfloat}{
	\parindent \z@
	\leftskip \z@ \@plus 1fil \@minus \textwidth
	\rightskip\leftskip
	\parfillskip \z@skip}
\makeatother

\begin{document}
\title{Fourier analysis of a delayed Rulkov neuron network}
\date{\today}
\author{Roberto Lozano}
\email[Corresponding author ]{roberto.lozano@urjc.es}
\affiliation{Nonlinear Dynamics, Chaos and Complex Systems Group,
Departamento de F\'{i}sica, Universidad Rey Juan Carlos, Tulip\'{a}n
s/n, 28933 M\'{o}stoles, Madrid, Spain}
\author{Javier Used}
\affiliation{Nonlinear Dynamics, Chaos and Complex Systems Group,
Departamento de F\'{i}sica, Universidad Rey Juan Carlos, Tulip\'{a}n
s/n, 28933 M\'{o}stoles, Madrid, Spain}
\author{Miguel A.F. Sanju\'{a}n}
\affiliation{Nonlinear Dynamics, Chaos and Complex Systems Group,
Departamento de F\'{i}sica, Universidad Rey Juan Carlos, Tulip\'{a}n
s/n, 28933 M\'{o}stoles, Madrid, Spain}
\affiliation{Department of Applied Informatics, Kaunas University of Technology, Studentu 50-415, Kaunas LT-51368, Lithuania}

\date{\today}

\begin{abstract}
	
	We have analyzed the synchronization of a small-world network of chaotic Rulkov neurons with an electrical coupling that contains a delay. We have developed an algorithm to compute a certain delay whose result is to improve the synchronization of the network when it was slightly synchronized, or to get synchronized when it was desynchronized. Our general approach has been to use tools from signal analysis, such as Fourier and wavelet transforms. With these tools, we have characterized the behavior of the neurons for different parameters in frequency and time-frequency domains. Finally, the robustness of the algorithm has been tested by using non-homogeneous neurons affected with a parametric noise.
	
	
\end{abstract}

\maketitle

\section{Introduction}

		Neuronal dynamics constitutes an important field of research in mathematical biology and biophysics. This research increases the understanding of mental illnesses, provides tools for their diagnosis, or simply helps to understand fundamental processes in the brain.
		
		Many aspects of the dynamics of the brain can be described by using mathematical models, which basically can be divided in continuous (ODEs or PDEs) or discrete in time (Recurrence equations or maps). Each one of them possesses advantages and disadvantages. Continuous models have a higher computational cost, which is increased even more when we consider them forming a network. Some well-known examples are the Hodgkin–Huxley, the FitzHugh–Nagumo, the Hindmarsh–Rose and the Kuramoto models \cite{Huxley,Nagumo,Hindmarsh,Kuramoto}. In spite of the possible limitations of discrete-time models, they possess a clear advantage from the computational point of view. Since they are basically constituted by recurrence relations, they can be easily computed, so that the computation time is much lower than in the case of continuous models. Furthermore, the mathematical analysis can be simpler. However, the interpretation of the results might not be that simple. Among the discrete-time models (also called map-based neuron models) we can mention the Izhikevich, the non-chaotic Rulkov, and the chaotic Rulkov models, among others \cite{Izhikevich,nonRulkov,Rulkov,ibarz2011}. 
		
		In this paper, we want to analyze a more complex structure where neurons are connected between them forming a particular network. Consequently, the mathematical tools that represent a certain pattern of connections are graphs, where the nodes represent the neurons, and the edges the connections. They can be expressed in matrix notation by using an adjacency matrix, where the neurons $i$ and $j$ are connected if the entry $(i,j)$ is 1,  while they are not connected if entry $(i,j)$ is 0.
		
		Small-world networks have been used in order to reproduce the brain connectivity \cite{Brain}. They are a kind of random graphs where the typical distance between nodes $L$ scales as the logarithm of the number of nodes $N$, i.e., $L\propto\log{N}$. Among other relevant properties we can mention that some nodes possess much more connections than the average node, and their nodes usually form clusters \cite{Watts1998}.	 
		
		We are considering the case in which the interaction between two neurons is modeled mathematically with an electrical coupling \cite{Nordenfelt}, that is, we are not considering for convenience chemical couplings. Physiologically, this represents the transportation of ions through the gap junction channel between very close neurons. Moreover, the electrical couplings have been implemented with a time delay representing the communication time between neurons.
		
		Finally, we have also assumed that not all neurons are equal, i.e., they do not have the same values of their parameters. We model this non-homogeneity by adding a Gaussian white noise in a key parameter of the chaotic Rulkov neuron.
		
		The synchronization of the neurons in a network is a very interesting emergent property. One of the practical interest of this property is its relationship with some mental diseases. In particular, in the Alzheimer disease or schizophrenia, the synchronization of neurons is lower than the expected. However, in epilepsy or Parkinson disease the synchronization of neurons is greater than expected \cite{neuralill,neuralill2}.
		
		Our approach is based on signal analysis, i.e., using Fourier and wavelet transform. The point of these tools is that the Fourier transform allows us to study the frequency domain of our data, while wavelet transforms do the same but in a time-frequency domain. With the Fourier transform we can study the frequencies in an asymptotic stable oscillation while by using wavelet transforms, we can study the evolution of these frequencies in time. 
		
		The main goal of this paper has been to develop an algorithm that improves the synchronization of neuron network. This algorithm computes a delay that included into the electrical coupling, helps to improve the synchronization of the neuron network. We tested this algorithm for a non-homogeneous neuron network, showing its robustness for low and medium noise intensities.

\section{The Model}
		
		
		The chaotic Rulkov model \cite{Rulkov,ibarz2011} is a well-known map-based neuron model defined as
		\begin{equation}
		\label{eq:Rulkov}
			\begin{aligned}
				&x_n=\frac{\alpha}{1+x^2_{n-1}}+y_{n-1}, \\
				&y_n=y_{n-1}-\beta\left(x_{n-1}-\sigma\right),
			\end{aligned}
		\end{equation}
		where $\alpha$, $\beta$ and $\sigma$ are constant parameters, the variable $x_n$ is the fast variable representing the transmembrane voltage of a single neuron, and $y_n$ is the slow variable standing for the slow gating process. The difference of the time scales between the two variables is determined by a sufficiently small value of the parameter $\beta$, i.e., $0< \beta \ll 1$. For different choices of the parameters, the model described in Eq.~\eqref{eq:Rulkov} can show different behaviors of the neurons in a qualitative manner, such as: rest, spikes, bursts, and chaotic spikes \cite{Parameter}.	
		
		As we explained earlier in the Introduction, the main goal of this paper is to study the global behavior of some coupled neurons, as it happens in the real brain. For this purpose, we need to connect these neurons mathematically using a network, in our case we will use a small-world network. 
		
		If we want to build a small-world network, we need to order the $N$ nodes, and then connect each node with its $k$ first neighbors in a cyclical way. For instance, the first node should be connected with the $k$ latest nodes. Finally, each node has a probability $p$ to rewire one of these edges with any other node. Thus, this kind of graph has three parameters: (i) $N$, the number of nodes; (ii) $k$, the number of first neighbors connected with one neuron; and (iii) $p$, the probability of rewiring an edge. This last parameter indicates the random nature of the graph, i.e., as $p$ increases the graph will be more random.	
		
		The randomness of the graph implies that, for the same choice of parameters, there might be different degrees of synchronization that will depend on the graph itself. This phenomenon makes sense in the real brain, since all brains have the same macroscopic structure but from a microscopic point of view the connection patterns are different for every person.
		
	
		The node connections given by the graph represent electrical couplings between the neurons. Mathematically, the coupling is modeled with the addition of a term in our model equation. Furthermore, since the distance between neurons is not zero, we have a non-zero electrical transmission time, what implies that we need to introduce a delay $\tau$ in the coupling. For the neuron $i$, the mathematical representation of these couplings have the form:
		\begin{equation}
			\label{eq:electricalcoupling}
			\sum_j\delta_{ij}(x_{j,n-\tau}-x_{i,n-1}),
		\end{equation}
		where $i$ is the evaluated neuron, $j$ the remaining neurons, and $\delta_{ij}$ a coupling constant between neurons $i$ and $j$. Since the electrical connection between two neurons is bidirectional, the coupling strength $\delta_{ij}$ is equal to $\delta_{ji}$. This coupling would be equal to zero if and only if $i$ and $j$ are not connected.

		
		Summarizing all this previous information, we proceed to build the complete model of the network. In this case we will also fix the same value for the delay and the coupling strength for all neurons, so that the model can be written, in matrix notation, as
		\begin{equation}
		\label{eq:NetworkRulkov}
			\begin{aligned}
				&x_n=\frac{\alpha}{1+x_{n-1}^2}+y_{n-1}+ \\
				&\quad+\delta(Ax_{n-\tau}-Dx_{n-1}), \\
				&y_n=y_{n-1}-\beta(x_{n-1}-\sigma),
			\end{aligned}
		\end{equation}
		where $x$ and $y$ are vectors, and the entry $i$ is the value of $x$ or $y$ for the neuron $i$; $A$ is the adjacency matrix of the graph, $\tau$ the delay, and $\delta$ is a coupling constant defined as $\delta=1/(3(k+1))$. The operator $D$ is a diagonal matrix where each entry $d_{ii}=\sum_ja_{ij}$, i.e., it is the number of connections of the neuron $i$. The operations $x^2$ and $\alpha/(1+x^2)$ are point-wise for each element of the array. The coupling term is the same as in Eq.~\eqref{eq:electricalcoupling} but in a matrix notation:
		\begin{equation*}
			\begin{aligned}
				&\sum_j\delta_{ij}(x_{j,n-\tau}-x_{i,n-1})= \\ 
				&=\delta\left(\sum_ja_{ij}(x_{j,n-\tau}-x_{i.n-1})\right)= \\
				&=\delta\left(\sum_j(a_{ij}x_{j,n-\tau})-d_{ii}x_{i,n-1}\right).
			\end{aligned}
		\end{equation*}
		The first term is the matrix vector product $Ax_{n-\tau}$ and the second term is $Dx_{n-1}$.
		
		Some properties that we can study about the global behavior of the network are periodic; in fact, we will focus on them. These properties can be easily studied using spectral methods like the Fourier or the wavelet transform. So, once the time signals for every neuron have been obtained, we use the FFT (Fast Fourier Transform) and the CWT (Continuous Wavelet Transform) algorithms of MATLAB for their analysis. 
		
		The FFT gives us information about the phase and the frequency of the neurons while the CWT performs a time-frequency analysis, where we can see how the frequency varies in function of time. We use the Morse wavelet with symmetry parameter $\gamma=3$ and time-bandwidth product $P^2=60$ \cite{Olhede2002}.

\section{Frequency Analysis for a Single Neuron}

	In this section we will analyze the behavior of the Rulkov neurons (Eq.~\eqref{eq:Rulkov}) for different values of the parameter $\alpha$. In our case, we will analyze not only the time series but its frequency domain, because it will give us a better understanding of the behavior of the system.
	
\begin{figure*}
		\centerfloat
		\includegraphics[width=1.15\textwidth]{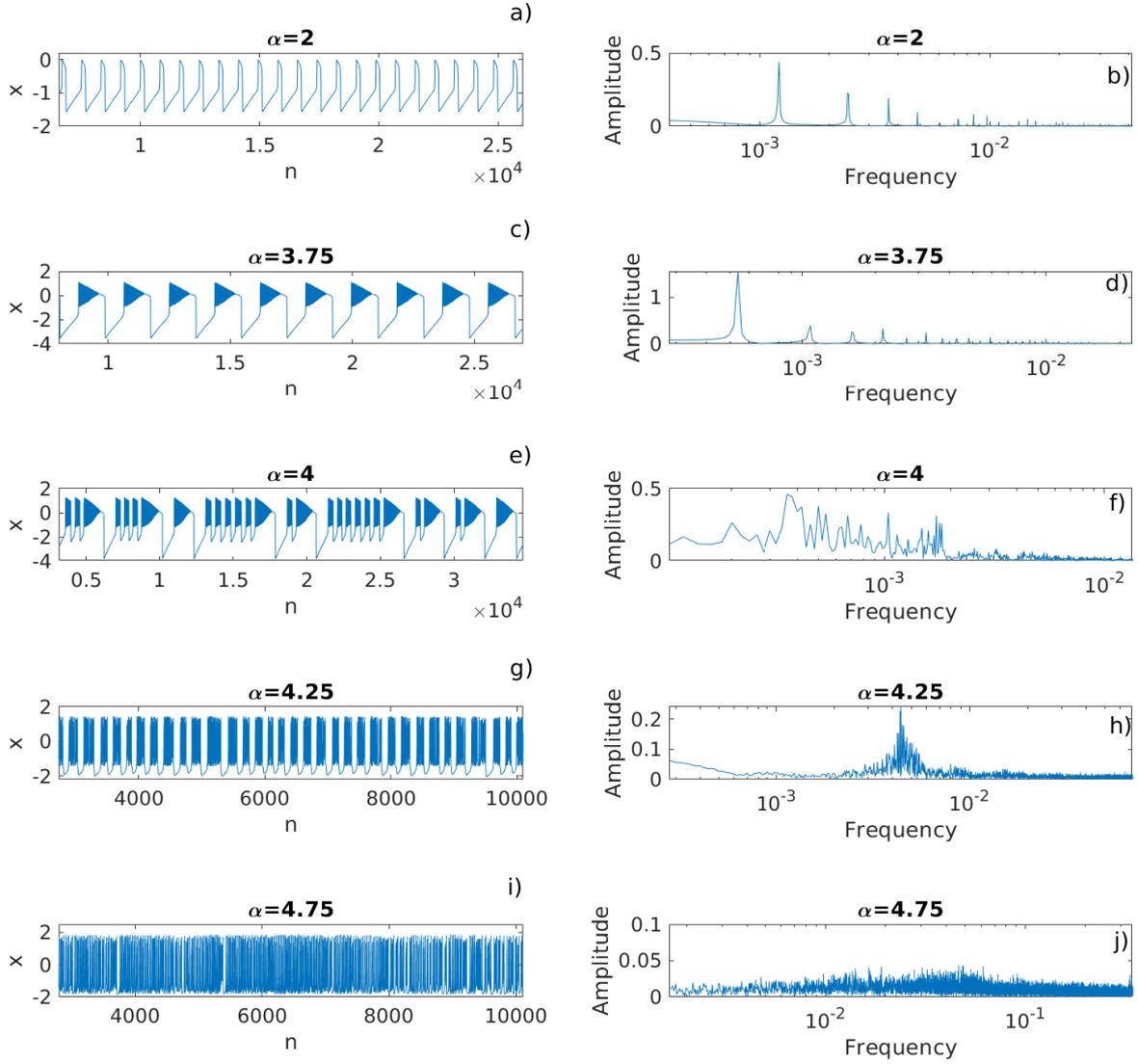}
		\caption{\textbf{Times series (left panels) and their corresponding Fourier transform (right panels) of one chaotic Rulkov neuron for different values of the parameter $\alpha$}. Parameters $\beta=10^{-3}$ and $\sigma=-1$ are fixed. In (a) we can clearly observe spikes that are transformed into a fundamental frequency and its harmonics in frequency domain in (b). In (c), we see long spikes with a decreasing oscillation after their maxima. The transform of (c) is represented in (d), which is similar to (b), where the maxima are its fundamental frequency and its harmonics. In (e), there are different combinations of long spikes and bursts. This is translated in one maximum for each combination in (f). In (g), we observe bursts that have a noisy spectrum centered in a fundamental frequency, as shown in (h). Finally, the chaotic bursts observed in (i) have a noisy spectrum in (j).}
		\label{fig:1neurona}
	\end{figure*}

	We present some of the more interesting results in Fig.~\ref{fig:1neurona}. On the left panels we represent the time series corresponding to different values of $\alpha$, while in the right panels we plot their corresponding Fourier transform. We can observe that the time series present the typical spiking behavior for low values of $\alpha$,  where their corresponding Fourier transforms have a maximum for their fundamental frequency and its harmonics, as shown in Figs.~\ref{fig:1neurona}(a-b). As $\alpha$ increases, the previous spiking regime starts to present a decreasing oscillation after having reached its maximum. In this case, we obtain a transform quite similar to the one shown for lower values of $\alpha$. Both cases present the characteristic spectrum of periodic but not smooth functions like the sawtooth function, as shown in Fig.~\ref{fig:1neurona}(c-d).
	
	If we increase a bit more the value of $\alpha$, we reach values where the behavior changes dramatically, presenting a mixture of spikes interspersed with a variable number of bursts. For this case, the frequency domain shows different maxima corresponding to each kind of bursts-spike group, as shown in Fig.~\ref{fig:1neurona}e-f.
	
	Finally, when the value of $\alpha$ is higher the system presents a plain bursting dynamics, as can be observed in Fig.~\ref{fig:1neurona}g. Its corresponding spectrum is centered in the principal firing frequency that can be seen in Fig.~\ref{fig:1neurona}h. The lower maxima in frequencies represents the internal oscillations inside the bursts. Finally, if we increase a bit more the value of $\alpha$, we have chaotic bursts characterized by a noisy spectrum, where no maxima are observed in the frequency domain, as illustrated in Fig.~\ref{fig:1neurona}(i-j).
	
\begin{figure*}
		\centerfloat
		\includegraphics[width=0.52\linewidth]{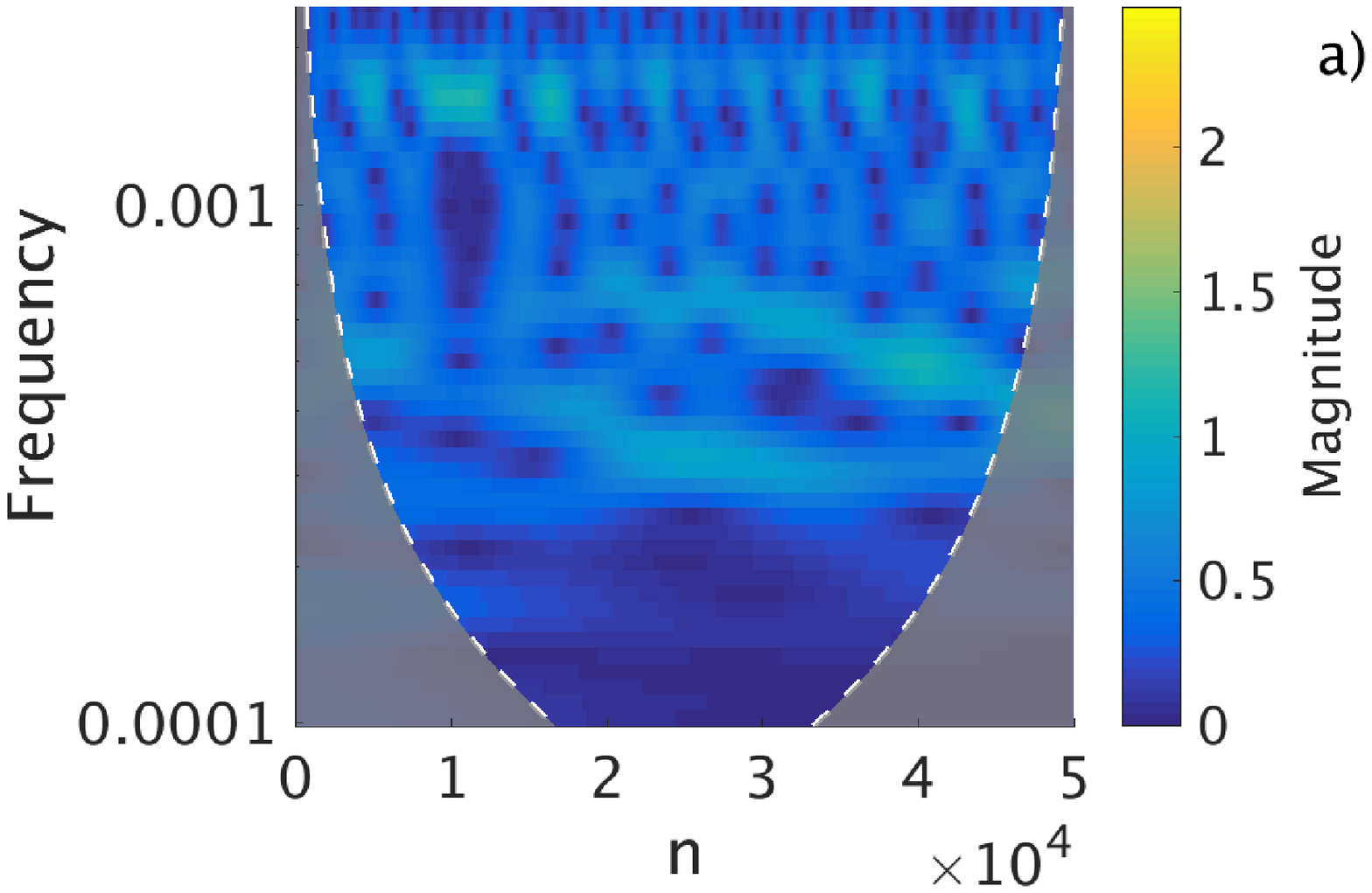}
		\includegraphics[width=0.52\linewidth]{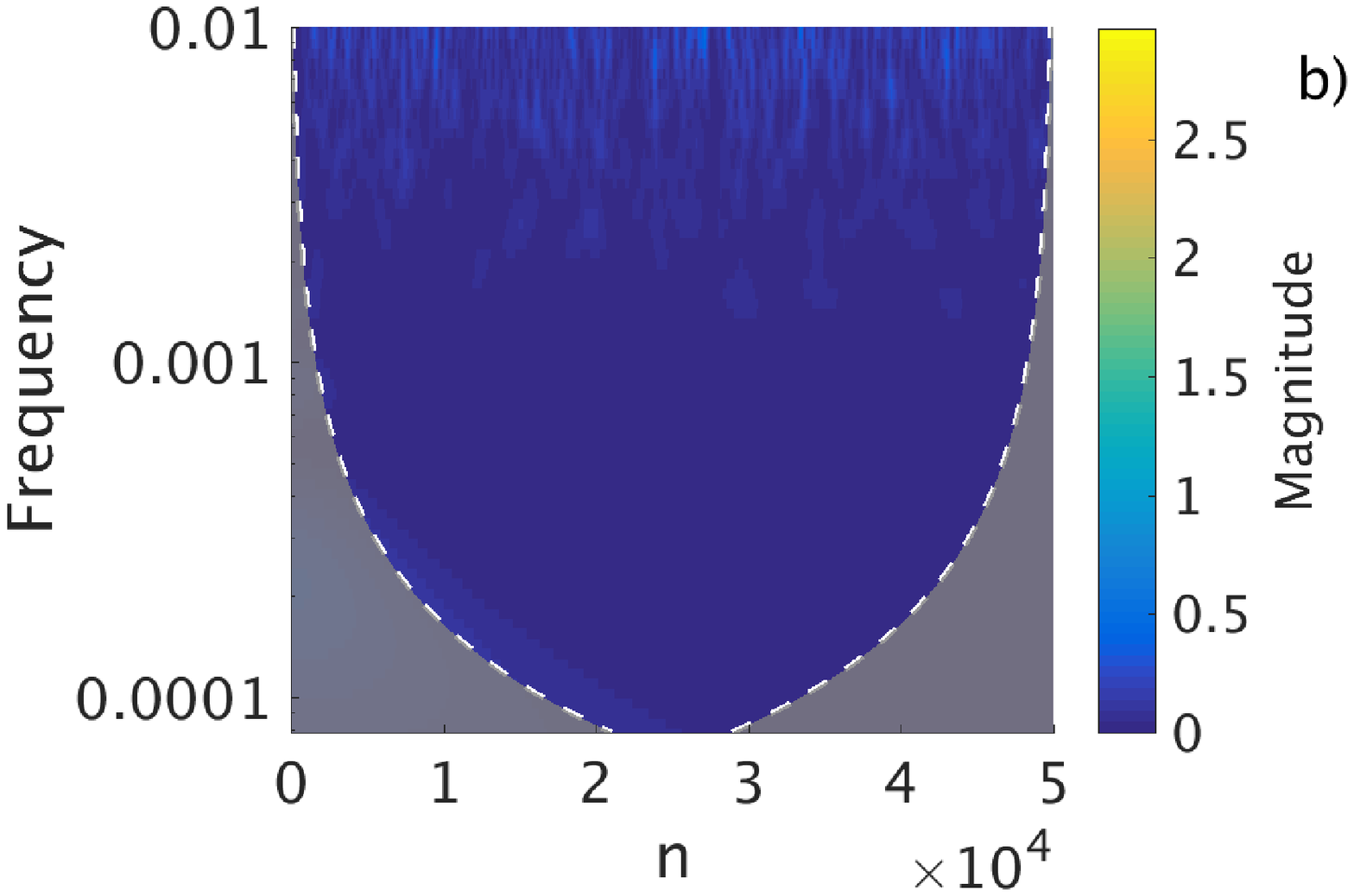}		
		\caption{\textbf{Wavelet transform of the time series of the Rulkov neuron for $\alpha=4$, (a), and $\alpha=4.75$, (b)}. The points in the grey region are not reliable due to the characteristics of the method, so that they have no interest. In (a), there is a region of frequencies varying in magnitude over time, which symbolizes the change in the number of bursts before the spikes. In (b), since we have chaotic oscillations, we have a broadband spectrum and consequently we cannot observe any fundamental frequency.}
		\label{fig:cwtalphas}
	\end{figure*}

	In the cases $\alpha=4$ and $\alpha=4.75$, we have observed that the neurons did not stabilize in a specific periodic oscillation; so we have used the wavelet transform for a better understanding of these neurons in their time-frequency behavior. In the case $\alpha=4$ there might be a periodic pattern of bursts-spike groups. And in the case $\alpha=4.75$ there might be some kind of order in the chaotic oscillations. Our numerical simulations shown in Fig.~\ref{fig:cwtalphas}a indicate that there is a kind of noisy alternation of the frequencies implying a change of the number of bursts for each kind of bursts-spike group, as we expected. At the same time, no fundamental frequency appears at any time as shown in Fig.~\ref{fig:cwtalphas}b. This kind of behavior is expected from a chaotic oscillation.
	
\section{Neurons in a Small-World Network}	

	When we have a group of interacting neurons, a very interesting global property is their possible synchronization. In the rest of the paper we will focus on the study of this property. We use a use a small-world network with parameters $N=50$, $p=0.2$ and $k=2$ to model the connections of the neurons.
	
\begin{figure*}
		\centerfloat
		\includegraphics[width=1.2\textwidth]{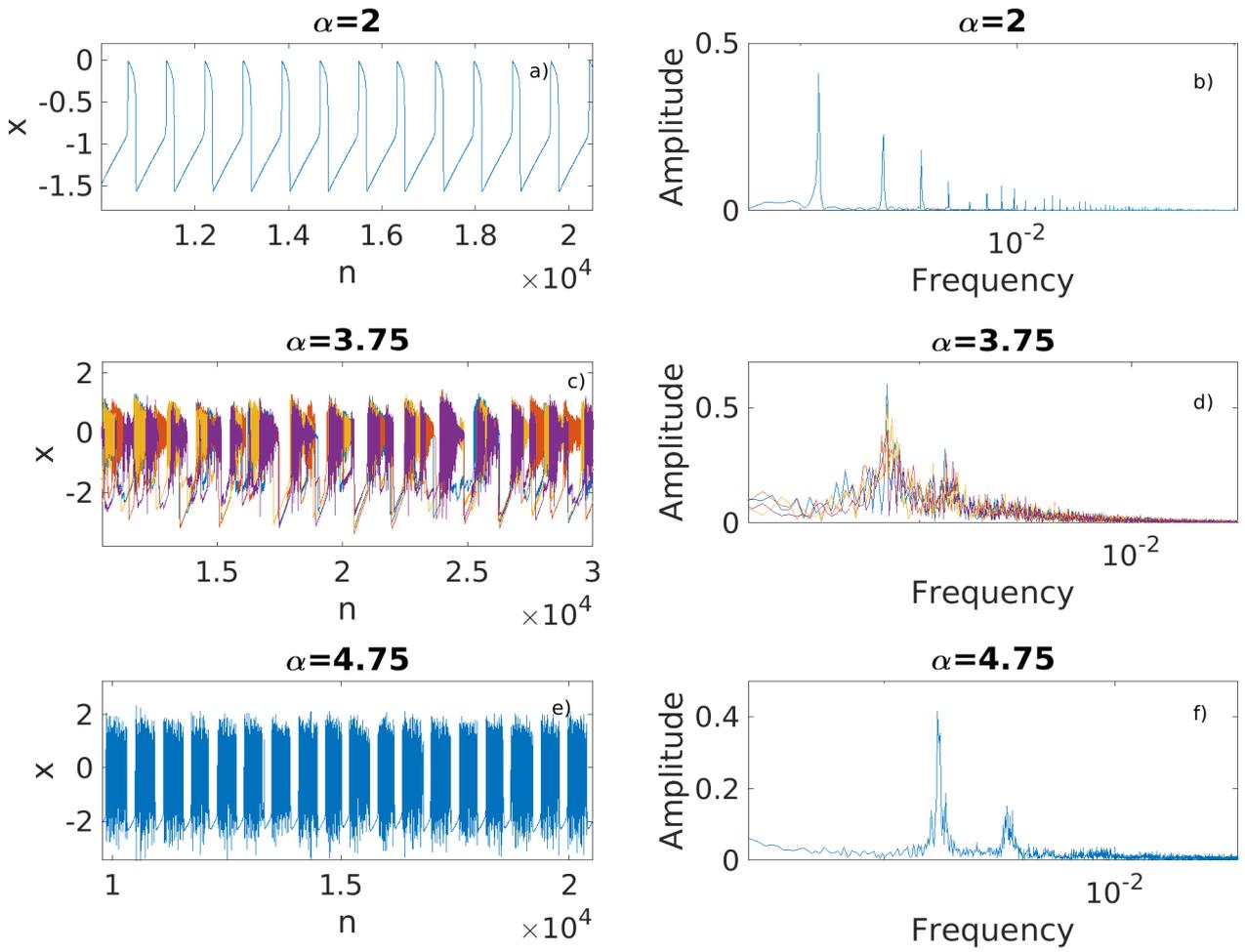}
		\caption{\textbf{Times series (left panels) and their corresponding Fourier transform (right panels) of a small-world network formed by $N=50$ chaotic Rulkov neurons for three values of the parameter $\alpha$}. Parameters $\beta=10^{-3}$ and $\sigma=-1$ are fixed and the graph was built with $k=2$ and $p=0.2$. In (a) all neurons follow a spiking dynamics, so that in (b) we can see maxima in the fundamental frequency and its harmonics. In (c) and (d) we represent several neurons since they have different frequencies, therefore they are not synchronized. Each color represents one of these frequencies. In (c) we see the oscillations of some neurons, which have no pattern though they are coupled. In (d) they have similar spectra, but not the same maxima, so that they do not fire at the same time. In (e) there is a bursting dynamics, and in (f) there are two maxima, where the largest one represents the firing frequency.}
		\label{fig:50neuronas}
	\end{figure*}

	The numerical simulations of the dynamics of this network are plotted in Fig.~\ref{fig:50neuronas}. As we did previously for the case of a single neuron, we will start our analysis with a low value of the parameter $\alpha$, and will study the changes in the global behavior as $\alpha$ increases.

	In Figs.~\ref{fig:50neuronas}(a-b), we plot the results for $\alpha=2$. In this case, we observe how all neurons are spiking in a fully synchronized state. This behavior is similar to the one observed in the case with only one neuron (Figs.~\ref{fig:1neurona}(a-b)). When we increase the value of the parameter $\alpha$ $(\alpha=3.75)$, we observe that the neurons are not synchronized. This is shown in Fig.~\ref{fig:50neuronas}c, where the time series corresponding to different neurons are plotted in different colors. This desynchronization manifests itself in the corresponding frequency domain (Fig.~\ref{fig:50neuronas}d), where a certain dispersion can be observed around the largest maximum.
	
	Finally, if we increase the value of the parameter $\alpha$ to $4.75$, we observe that the system possesses a bursting dynamics, as shown in Fig.~\ref{fig:50neuronas}e. If we represent the corresponding transform, Fig.~\ref{fig:50neuronas}f, we see two maxima where the largest one corresponds to the firing frequency. Now, we can appreciate a very interesting change in the behavior of the network. The bursting dynamics that happens for $\alpha=4.25$ (see Fig.~\ref{fig:50neuronas}f) in the case of the single neuron, appears here for $\alpha=4.75$.
	
\begin{figure}
		\centering
		\includegraphics[width=0.7\linewidth]{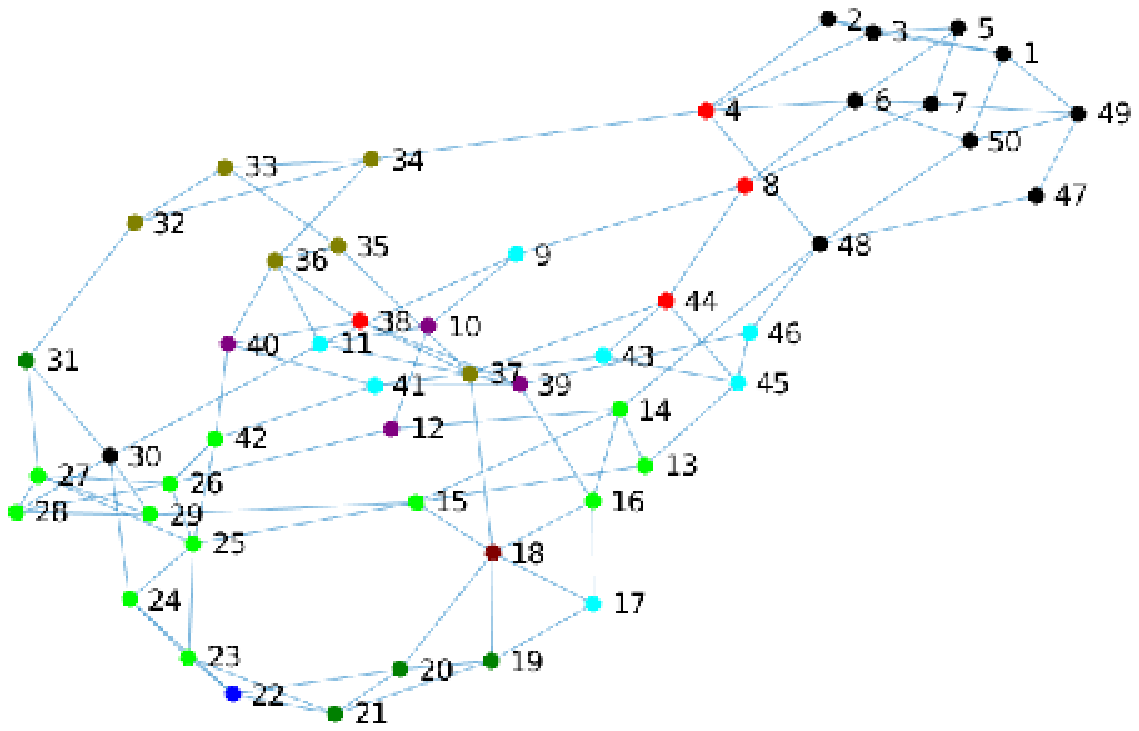}
		\caption{\textbf{Graph used in our simulation with a small-world network}.  Colors represent the different fundamental frequencies of each neuron for $\alpha=3.75$, illustrating that the neurons with similar fundamental frequencies are clustered.}
		\label{fig:Grapha375}
	\end{figure}

	The behavior of the neurons connected in a network depends of the network itself. Thus, their behavior or their synchronization will strongly depend on the corresponding network, even though their global behavior might be similar. The graph used in this section is plotted in Fig.~\ref{fig:Grapha375}, where each color represents a different fundamental frequency for the neuron in a simulation with $\alpha=3.75$. We can see that neurons with the same fundamental frequencies are usually clustered.

\section{Delay Algorithm}

	Analyzing the results obtained in the previous section, we observe that for some values of the parameter $\alpha$ not all neurons oscillate at the same frequency. One option for improving synchronization between the neurons is adding a delay to the model, so our goal here is to focus on developing an algorithm to calculate the delay $\tau$ for which this kind of systems can be synchronized showing its performance, as well.

	The steps of the algorithm that we have developed to calculate the delay $\tau$ used in Eq.~\eqref{eq:NetworkRulkov} are:
	\begin{enumerate}[(i)]
		\label{eq:algorithm}
		\item Compute the fundamental frequency (non-zero frequency with the largest amplitude) of each neuron.
		\item Calculate the period that corresponds to the fundamental frequency of the previous step.
		\item Use this period as the delay and compute again Eq.~\eqref{eq:NetworkRulkov}.
	\end{enumerate}
	
\begin{figure*}
		\centering
		\includegraphics[width=1\linewidth]{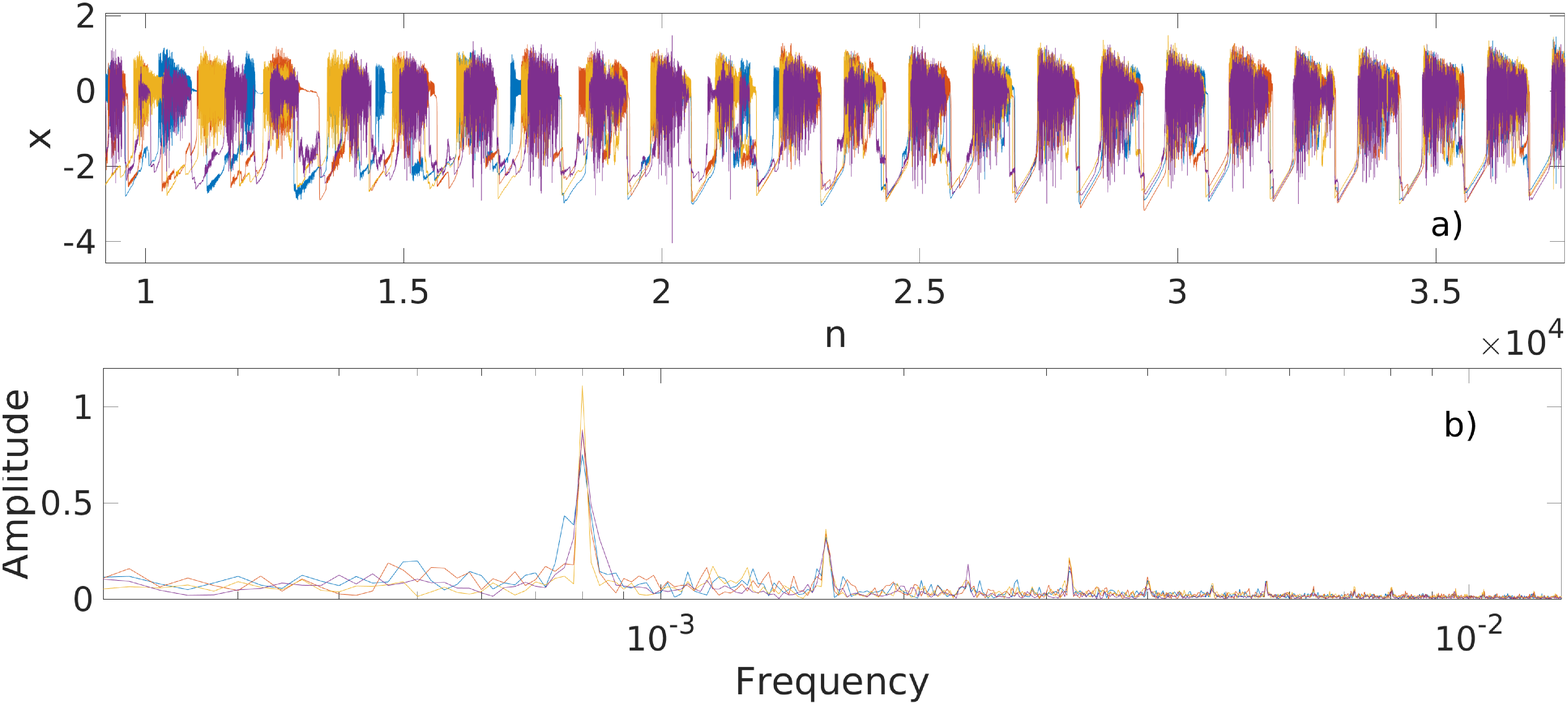}
		\caption{\textbf{Time series (up) and Fourier transform (down) for the same neurons and same parameters represented in Fig.~\ref{fig:50neuronas}c}. In (a) the neurons are synchronizing as $n$ grows, while in Fig.~\ref{fig:50neuronas}c they could not synchronize, (b) is similar to Fig.~\ref{fig:50neuronas}b, i.e., there is one fundamental frequency and its harmonics. This implies more order in the system, so that in Fig.~\ref{fig:50neuronas}d there is not a unique fundamental frequency, neither synchronization.}
		\label{fig:NWa375d}
	\end{figure*}

\begin{figure*}
		\centerfloat
		\includegraphics[width=1.2\linewidth,height=0.4\linewidth]{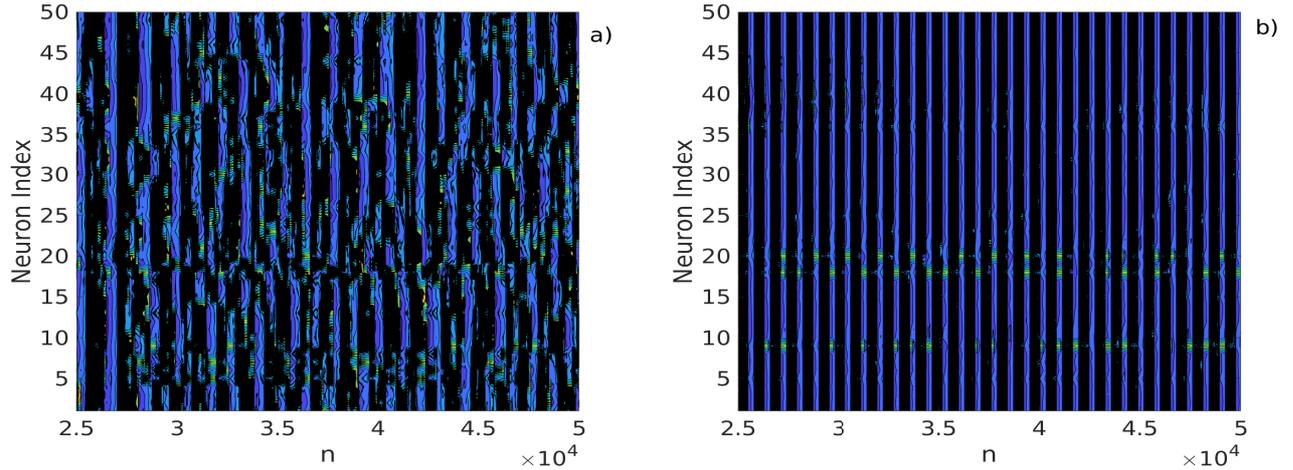}
		\caption{\textbf{Space-time plots of our small-network for $\alpha=3.75$}. In the left panel there is no delay, while in the right panel there is a delay. In (a) there is no synchronization, while in (b) the system is synchronized.}
		\label{fig:NoNoiseST}
	\end{figure*}

\begin{figure*}
		\centerfloat
		\includegraphics[width=0.52\linewidth]{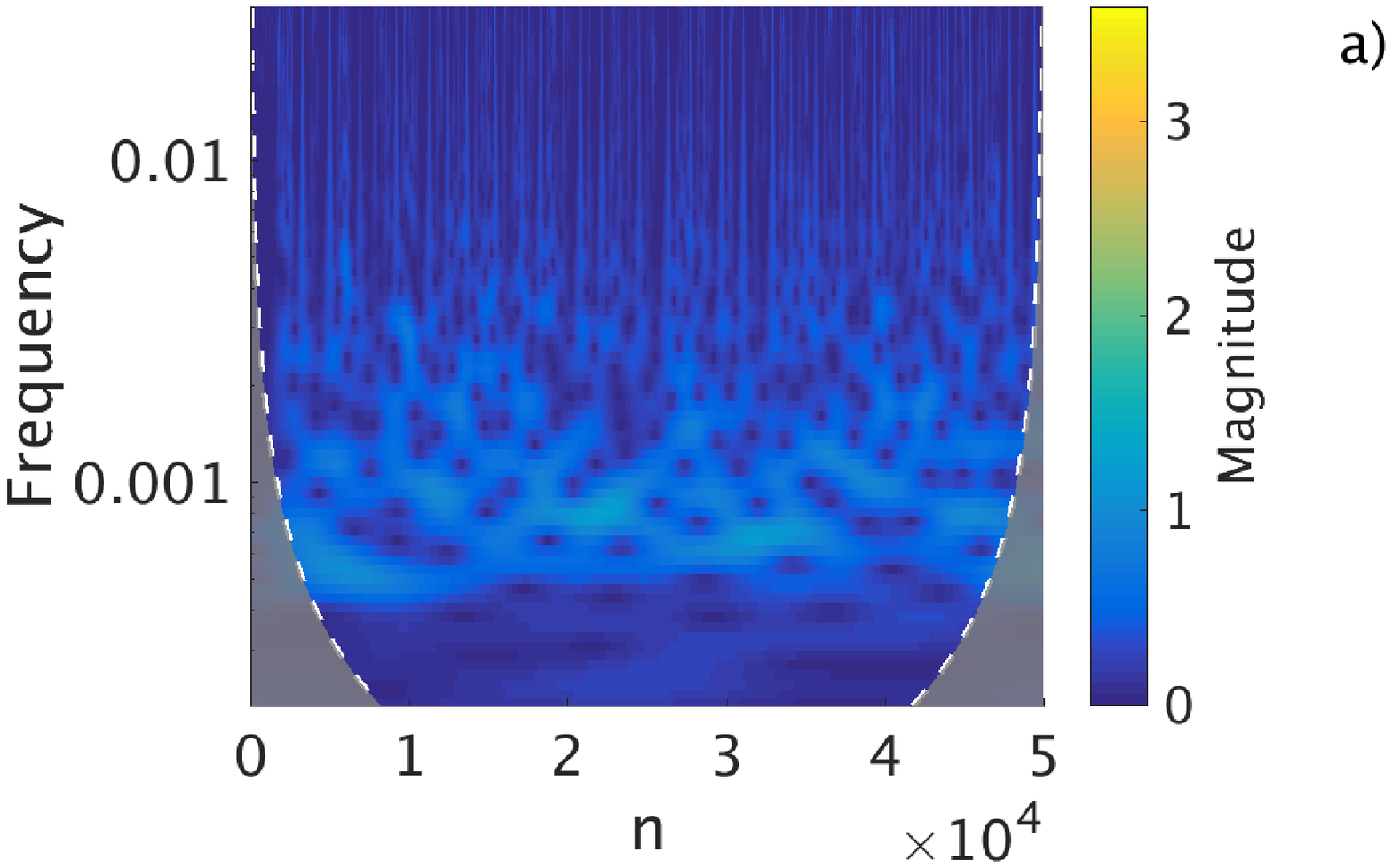}
		\includegraphics[width=0.52\linewidth]{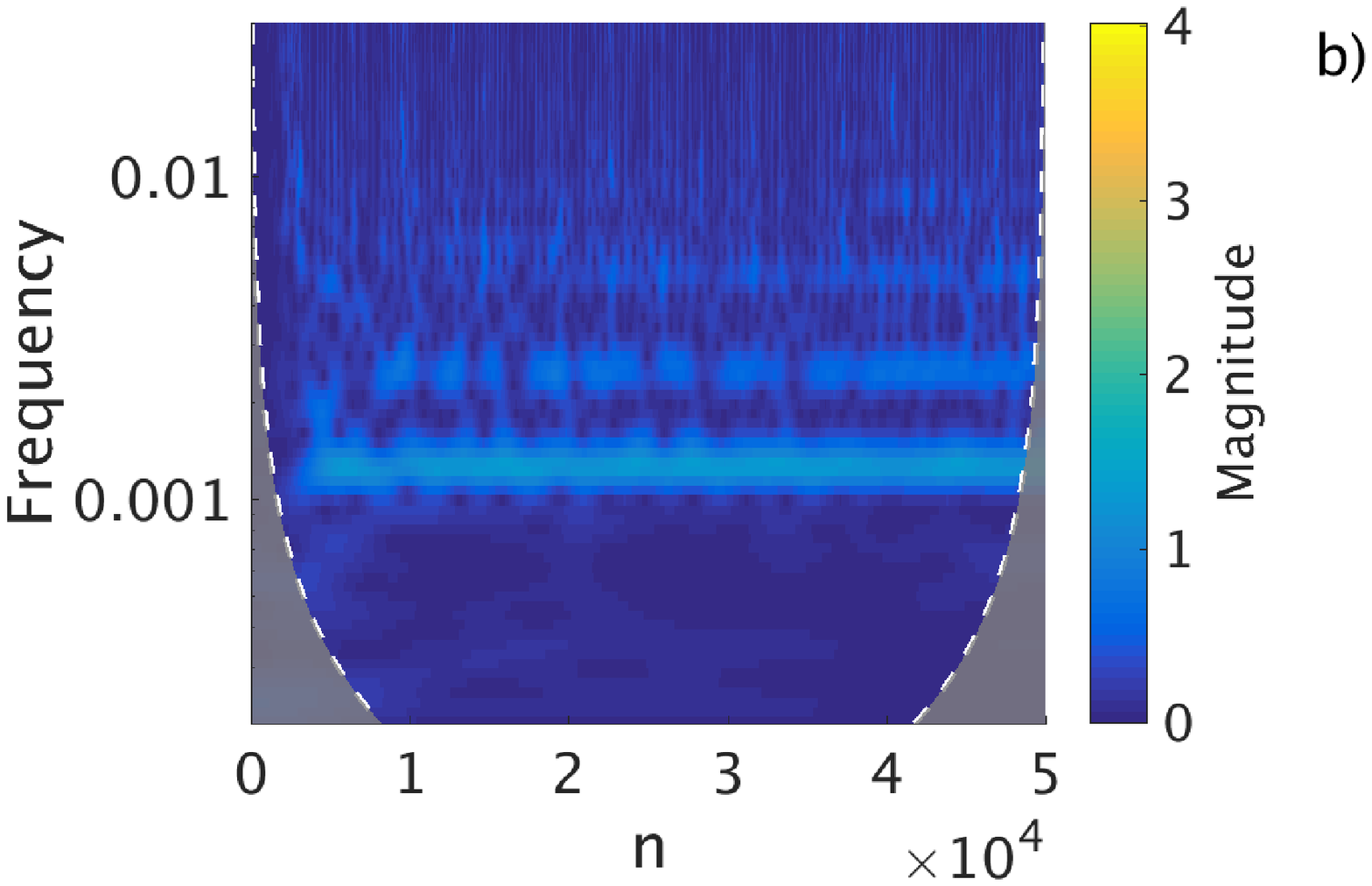}
		\caption{\textbf{Wavelet transforms for a random neuron extracted from the small-network for $\alpha=3.75$}. In the left panel there is no delay, while in the right panel there is a delay. The grey region contains points that are not reliable due to the characteristics of the method. In (a) there is no steady fundamental frequency. What actually occurs is an alternation of the fundamental frequencies. In (b), the delay induces a steady registry of frequencies.}
		\label{fig:50neuronasCWT}				
	\end{figure*}

	We use the delay $\tau$ obtained with this algorithm to solve the previous network for $\alpha=3.75$, using the same network parameters: $N=50$, $p=0.2$ and $k=2$. The results of this process are represented in Figs.~\ref{fig:NWa375d}, \ref{fig:NoNoiseST}b and \ref{fig:50neuronasCWT}b, in contrast with Figs.~\ref{fig:50neuronas}c and \ref{fig:50neuronas}d, and Figs.\ref{fig:NoNoiseST}a and \ref{fig:50neuronasCWT}a, where no delay was used in the simulation.
	
	In the time series of Fig.~\ref{fig:NWa375d} we plot the same neurons as represented in Fig.~\ref{fig:50neuronas}c. Now, we can observe that after an initial desynchronized state, the system starts synchronizing as time passes until its final state of synchronization. This synchronization is fast in frequency as we see in Fig.~\ref{fig:50neuronasCWT} but slow in phase, i.e., 30000 iterations are needed to achieve it. In the frequency domain plot, the maxima are harmonics of the fundamental frequency as in Fig.~\ref{fig:1neurona}b or Fig.~\ref{fig:50neuronas}b, in the same manner as what happens in a spiking dynamics. Space-time plots are very convenient to compare the synchronized and non-synchronized regimes, because they allow to show the state of all neurons for a certain time interval, helping to visualize whether the system is synchronized or not. 
		
	In Fig.~\ref{fig:NoNoiseST}, we draw the space-time plot of the latest iterations of the simulations. In the right panel where the delay has been used, the synchronization is clearly visible, while in the left panel, where there is no delay, there is no synchronization. In Fig.~\ref{fig:50neuronasCWT}, we have plotted the wavelet transforms of Fig.~\ref{fig:50neuronas}c and Fig.~\ref{fig:NWa375d}a. We can observe in the left panel, where we did not use any delay, that there is no steady frequency at all. However, we observe a steady fundamental frequency in the right panel where there is a delay. These results confirm all that we have already pointed out previously with the time series analysis.

\section{Effect of noise on $\alpha$ parameter}

	In order to simulate a more realistic neuron network, we are going to introduce some non-uniformity in the neurons. As a matter of fact, we are interested to analyze how this non-uniformity affects the global behavior of the network and the robustness of our delay algorithm.
	
	We implement this scheme with the addition of an internal noise in parameter $\alpha$. Specifically, we have used a Gaussian white noise $\xi$ with zero mean, variance one and uncorrelated in space, i.e., $\left<\xi_i,\xi_j\right>=2\delta(i-j)$, which is applied to the parameter $\alpha$ for each neuron. So that now we have $\alpha=\left[\alpha_1,\alpha_2,...,\alpha_n\right]^T $ in Eq.~\eqref{eq:NetworkRulkov}, where $\alpha_i=\alpha_0+D\xi_i$ and $D$ is the noise intensity. We fix $\alpha_0=3.75$ as in the previous section, and we use the same value of $D$ for all neurons.
	
\begin{figure*}
		\centerfloat
		\includegraphics[width=1.2\linewidth]{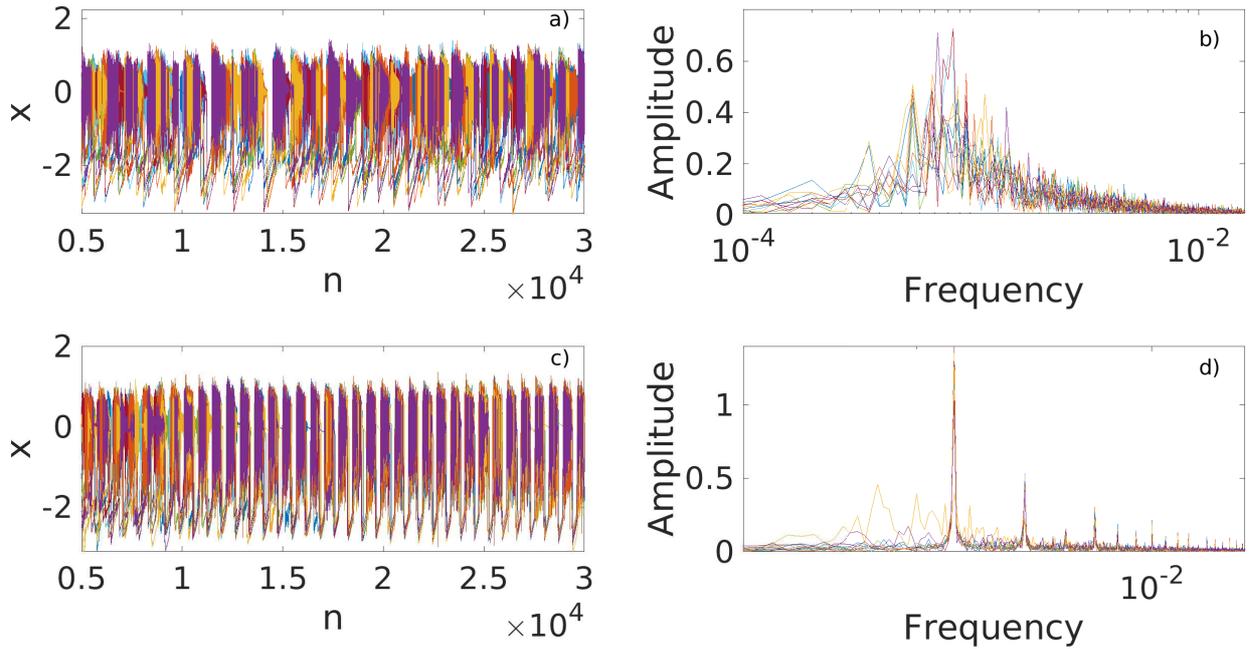}
		\caption{\textbf{Time series (left panels) and Fourier transform (right panels), for the small-world network with a low noise intensity in $\alpha$}. Panels (a-b) represent the dynamics without using the delay, while in panels (c-d) we use the delay. The parameter values are: $N=50$, $p=0.2$, $k=2$, $\alpha_0=3.75$ and $D=0.1$. Each color represents one fundamental frequency. In (a), there is no synchronization, and in (b) there is a great frequency dispersion. However the network can be synchronized by using the delay, as shown in (c) and (d).}
		\label{fig:LowNoise}
	\end{figure*}

\begin{figure*}
		\centerfloat
		\includegraphics[width=1.1\linewidth]{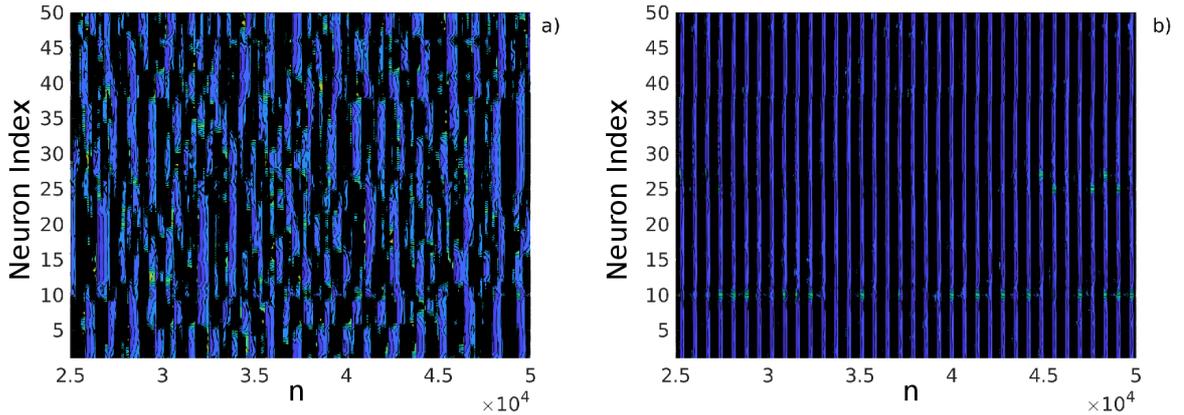}
		\caption{\textbf{Space-time plots of our small-world network obtained from our simulation with a low noise intensity in $\alpha$}. Left panel without delay and right panel with delay. In (a) there is no synchronization, but in (b) the network gets synchronized as if there were no noise.}
		\label{fig:LowNoiseST}
	\end{figure*}

	We have carried out several numerical simulations for different noise intensities, and without delay. For a low noise intensity, the system is slightly more desynchronized and the neurons present more fundamental frequencies than in the noiseless case (see Figs.~\ref{fig:LowNoise}(a-b)). Now, we proceed to repeat the simulation in presence of noise and with the delay. Our results show that the system gets synchronized due to the delay, as shown in Figs.~\ref{fig:LowNoise}(c-d). To compare those situations we have used their space-time plots, Fig.~\ref{fig:LowNoiseST}, where we can see the synchronizing effect of our delay.
	
\begin{figure*}
		\centerfloat
		\includegraphics[width=1.2\linewidth]{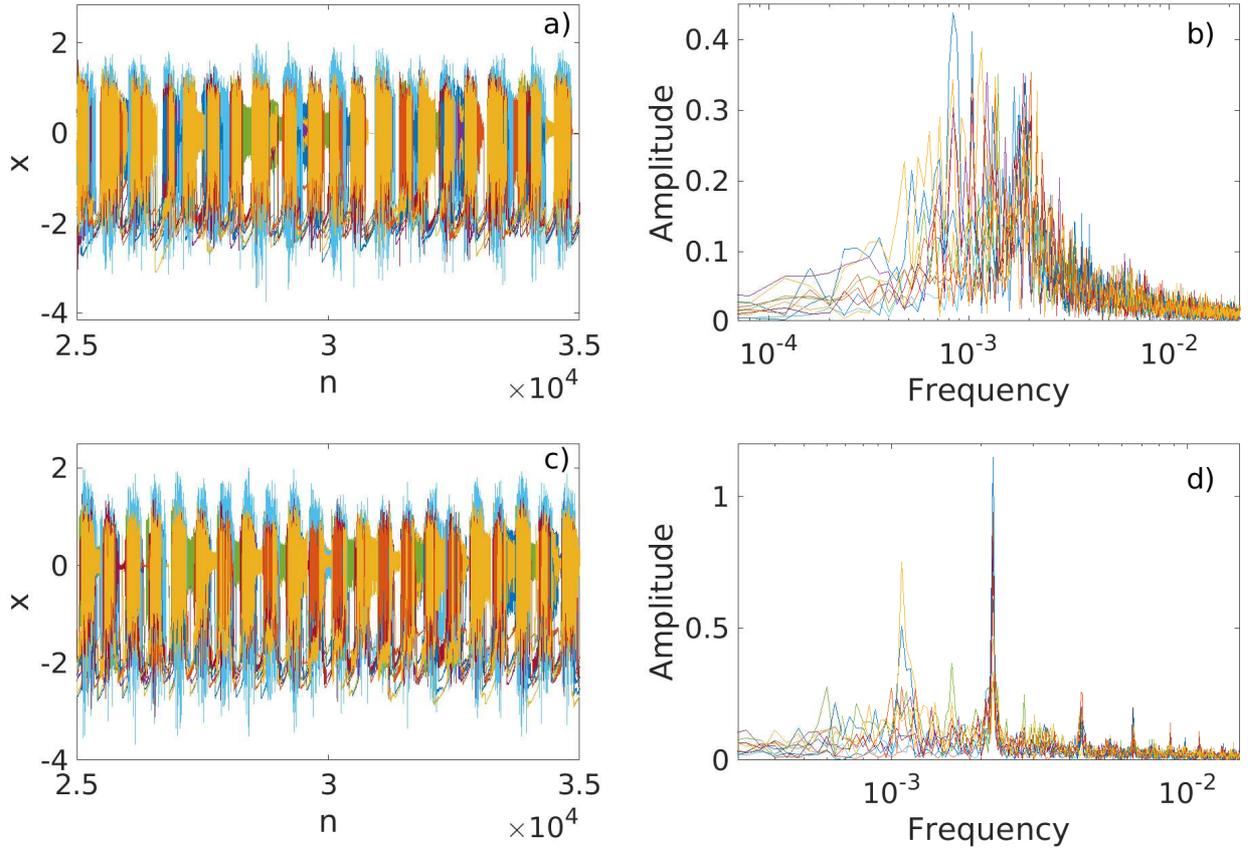}
		\caption{\textbf{Time series (left panels) and Fourier transform (right panels) for neurons in a small-world network with a high noise intensity in $\alpha$}. The upper panels were computed without using delay and the lower panels using the delay. The parameter values are: $n=50$, $p=0.2$, $k=2$, $\alpha_0=3.75$, and $D=0.75$. Each color represents one fundamental frequency. In this case the delay cannot synchronize all neurons.}
		\label{fig:HighNoise}
	\end{figure*}

\begin{figure*}
		\centerfloat
		\includegraphics[width=1.1\linewidth]{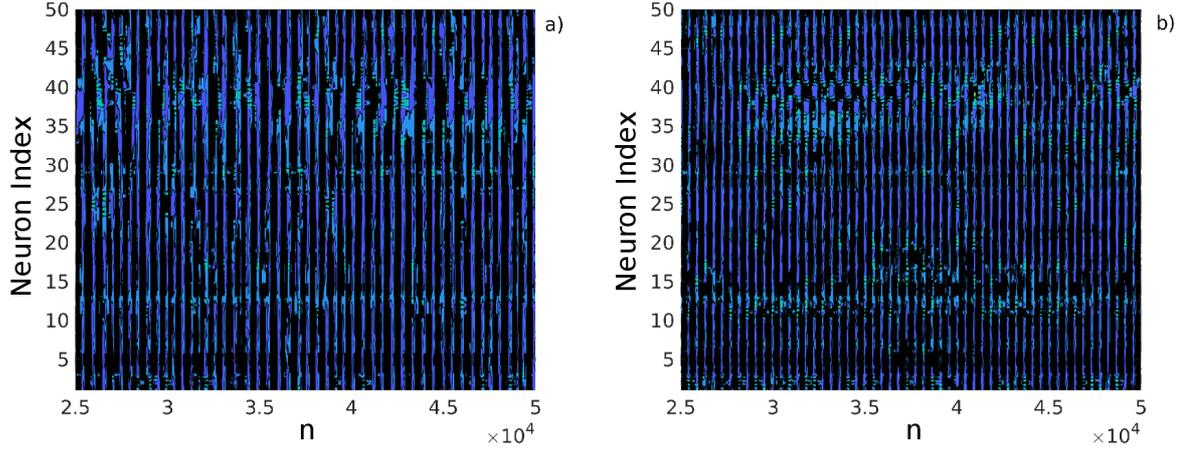}
		\caption{\textbf{Space-time plots of our small-world network obtained from our simulation with a high noise intensity in $\alpha$}. The left panel shows the case without delay and right panel the case with delay. Some sort of synchronization can be observed in (a), while in (b) the synchronization seems to have been improved.}
		\label{fig:HighNoiseST}
	\end{figure*}

	We are considering now a high noise intensity, where $D=0.75$. We use this value of $D$ because is the lower intensity for which the system cannot be synchronized using our delay in any of our simulations. With a high noise intensity, the system seems less desynchronized than the case with a low noise intensity (compare Fig.\ref{fig:LowNoise}a with Fig.\ref{fig:HighNoise}a, and Fig.\ref{fig:LowNoiseST}a with Fig.\ref{fig:HighNoiseST}a). This is counter-intuitive, since we would expect a more disordered state with a higher noise intensity.
	
	Even though the situation now seems more ordered, our algorithm cannot synchronize neurons completely. Yet it improves the synchronization, as it can be seen in Figs.~\ref{fig:HighNoise}(c-d) and \ref{fig:HighNoiseST}b. The reason is that the bursts do not need to rest before starting firing again. The fundamental frequency is the firing frequency, but here we have lower frequency maxima  representing bursts that do not rest between firing.
	
\begin{figure*}
		\centerfloat
		\includegraphics[width=0.52\linewidth]{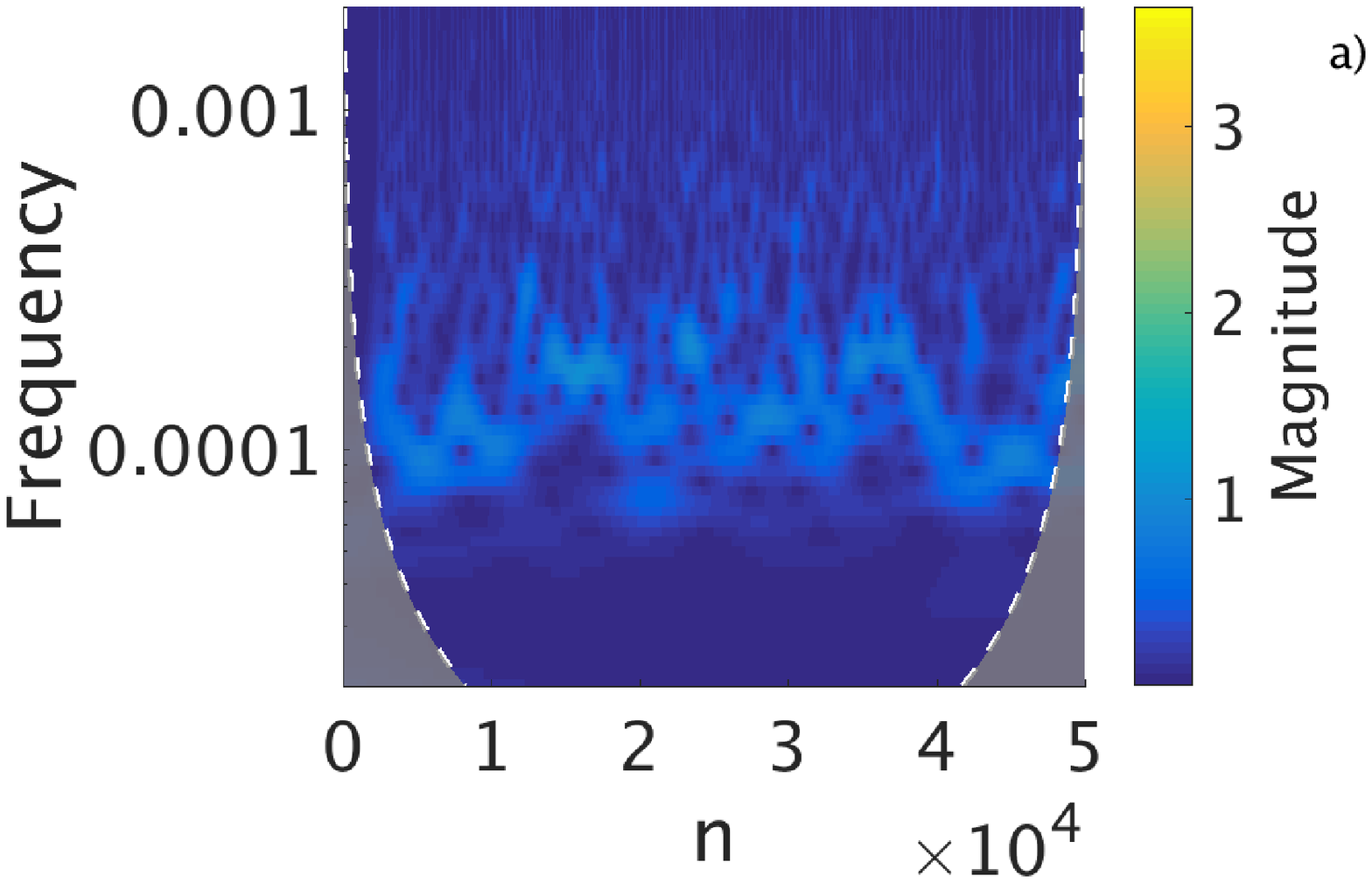}
		\includegraphics[width=0.52\linewidth]{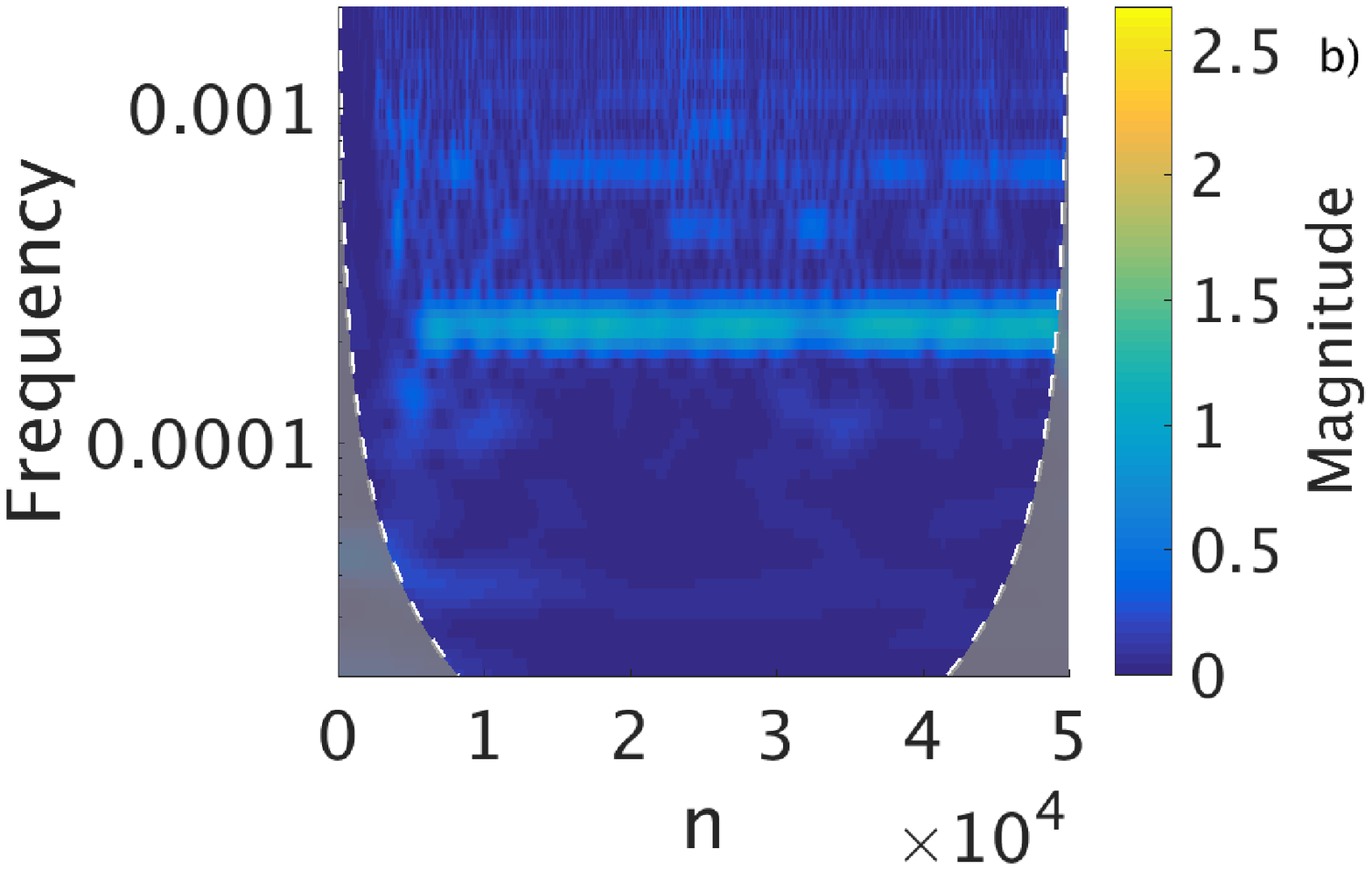}		
		\caption{\textbf{Wavelet transform for a random neuron extracted from our simulation with with a high noise intensity in $\alpha$}. As usual, the left panel is without delay and the right panel with it.  As in the previous cases, the grey region represents the points that are not reliable. In (a) we do not see any steady frequency, while in (b) we can see a steady fundamental frequency, though this does not happens for its harmonics.}
		\label{fig:cwtHighNoise}
	\end{figure*}

	In the left panel of the wavelet transform plot, Fig.~\ref{fig:cwtHighNoise}a, there is no fixed frequency over time. Nevertheless, in the right panel, Fig.~\ref{fig:cwtHighNoise}b, we can see a steady fundamental frequency but not in its harmonics. This could be an effect of the algorithm failure.

\section{Conclusion}
	
	We have characterized the behavior in frequency domain of chaotic Rulkov neurons. This provides useful information on the particular dynamics of the neurons. The same analysis has also been done on a small-world network of neurons. 
	
	We have been mainly focused in what happens with the synchronization in phase and frequency of the network by using tools derived from signal analysis. We have been interested in the effect of adding a delay to the network couplings. Our numerical simulations have proved that with the use of a particular delay, the synchronization of the system is improved when it was slightly synchronized or gets synchronized when it was desynchronized. We have developed an algorithm to compute this particular delay, which is related to the fundamental frequency of some neurons. 

	A more realistic approach to a real neuron network is to assume that not all neurons are equal, i.e., they do not have the same values of their parameters. To model this non-homogeneity with an internal white noise applied to parameter $\alpha$, which is one of the key parameters of the Rulkov neuron. We have tested our algorithm to compute the delay that synchronizes the network in this case. Our algorithm is robust against low and medium noise intensities, but it fails for higher intensities.

\section*{Acknowledgments}

	This work was supported by the Spanish State Research Agency (AEI) and the European Regional Development Fund (FEDER) under Project No.~FIS2016-76883-P.
		
	
\bibliography{Bibliografia}
	
\end{document}